\documentclass{article}
\usepackage[utf8]{inputenc}
\usepackage[english]{babel}
\usepackage[T1]{fontenc}
\usepackage{amssymb, graphicx, amsmath}
\DeclareMathOperator*{\argmin}{argmin}
\DeclareMathOperator*{\argmax}{argmax}

\title{Cross-validation of correlation networks using modular structure}
\author{Magnus Neuman$^1$, Viktor Jonsson$^{1,2}$, Joaquín Calatayud$^3$, Martin Rosvall$^1$}

\begin{document}
\date{}
\maketitle

\begin{small}
\noindent 1. Integrated Science Lab, Department of Physics, Umeå University, Umeå, Sweden\\
\noindent 2. School of Public Health and Community Medicine, University of Gothenburg, Gothenburg, Sweden\\
\noindent 3. Departamento de Biología, Geología, Física y Química inorgánica, Universidad Rey Juan Carlos, Madrid, Spain
\end{small}

\begin{abstract} 
Correlation networks derived from multivariate data appear in many applications across the sciences. These networks are usually dense and require sparsification to detect meaningful structure. However, current methods for sparsifying correlation networks struggle with balancing overfitting and underfitting. We propose a module-based cross-validation procedure to threshold these networks, making modular structure an integral part of the thresholding. 
We illustrate our approach using synthetic and real data and find that its ability to recover a planted partition has a step-like dependence on the number of data samples. The reward for sampling more varies non-linearly with the number of samples, with minimal gains after a critical point. A comparison with the well-established WGCNA method shows that our approach allows for revealing more modular structure in the data used here.

\end{abstract}

\section*{Introduction}

Many important applications deal with networks derived from correlations in multivariate data, such as gene co-expression networks \cite{Wang,Marbach}, 
fMRI scans of brain activity \cite{Bullmore} and ecological co-occurrence networks \cite{Barberan}. Correlation networks are dense, and finding an appropriate threshold to make the networks sparser and separate signal from noise is a critical first step in revealing meaningful structure. Researchers have developed a suite of methods to this end, including weighted gene co-expression network analysis (WGCNA) \cite{Horvath} that imposes soft thresholding to obtain a heuristically motivated scale-free network, hard thresholding \cite{Barberan, deVries} and consensus approaches \cite{Marbach}, while other researchers suggest that thresholding should be avoided \cite{Civier}. Methods not tied to a particular application include network coarse-graining and filtering techniques \cite{Serrano,Tumminello, Dianati}, and regularisation methods such as the graphical lasso \cite{Friedman} and neighbourhood selection \cite{Meinshausen}, which seek sparse estimates of the precision matrix. While these methods have particular advantages, they ignore whether the resulting pruned networks represent robust and informative structures. Since the overall objective when representing data as networks is often downstream analysis of network structure, existing thresholding approaches can limit our ability to reveal valuable patterns in the data. 

\begin{figure}[tb]
\includegraphics[scale=0.1]{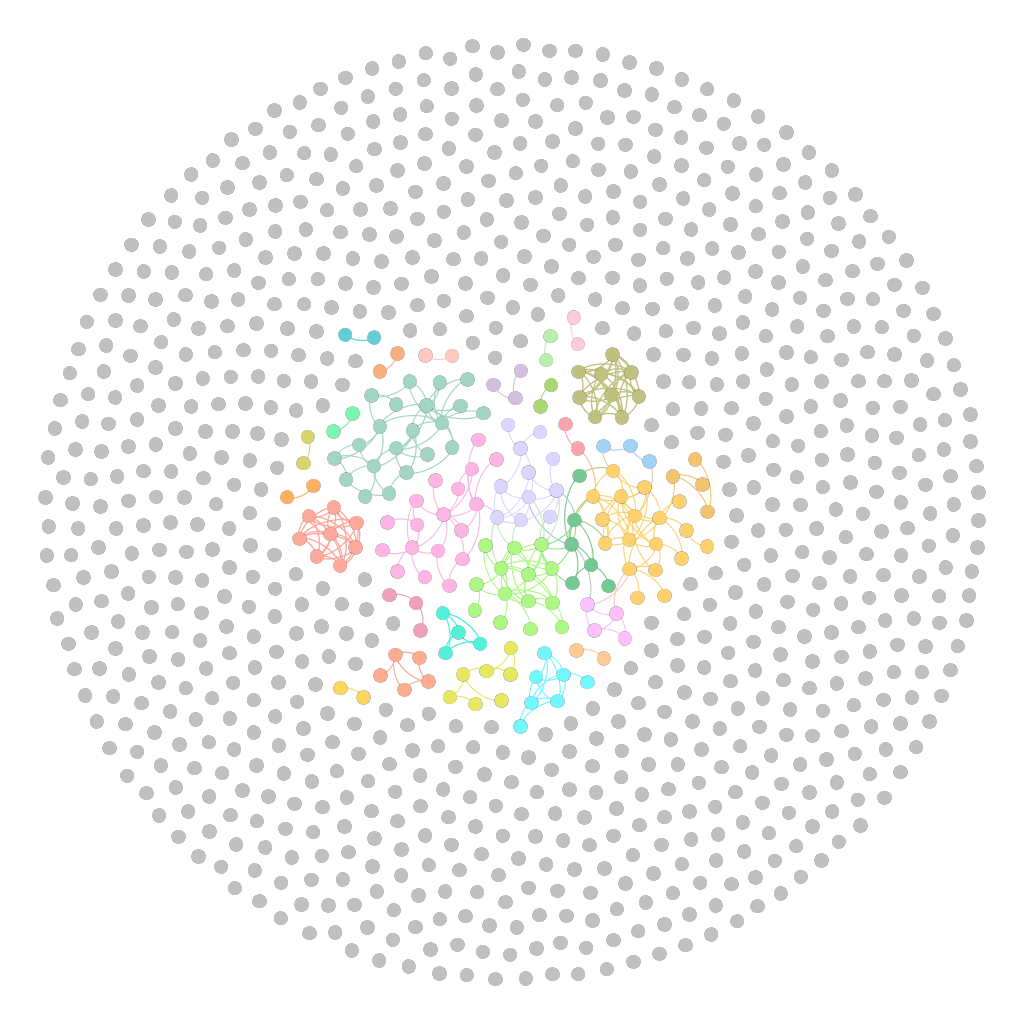}
\includegraphics[scale=0.1]{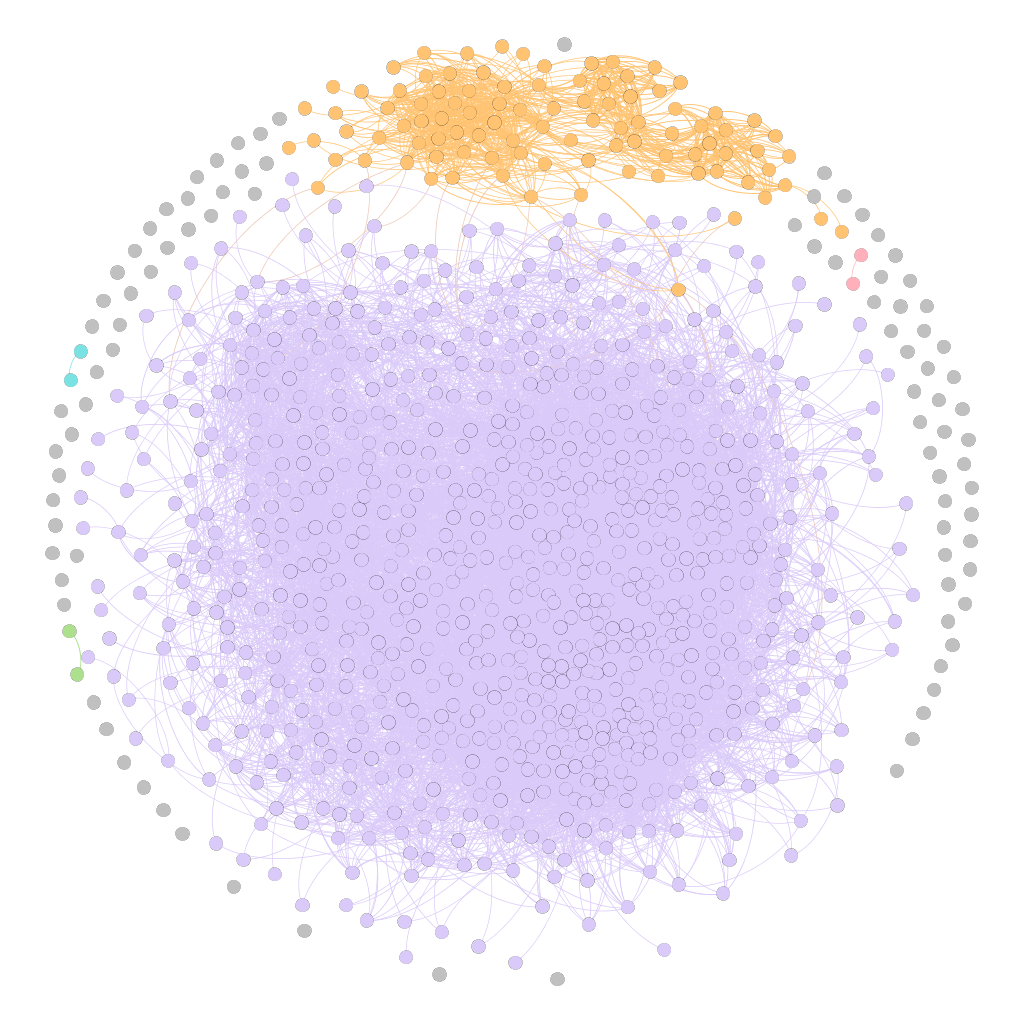}
\includegraphics[scale=0.1]{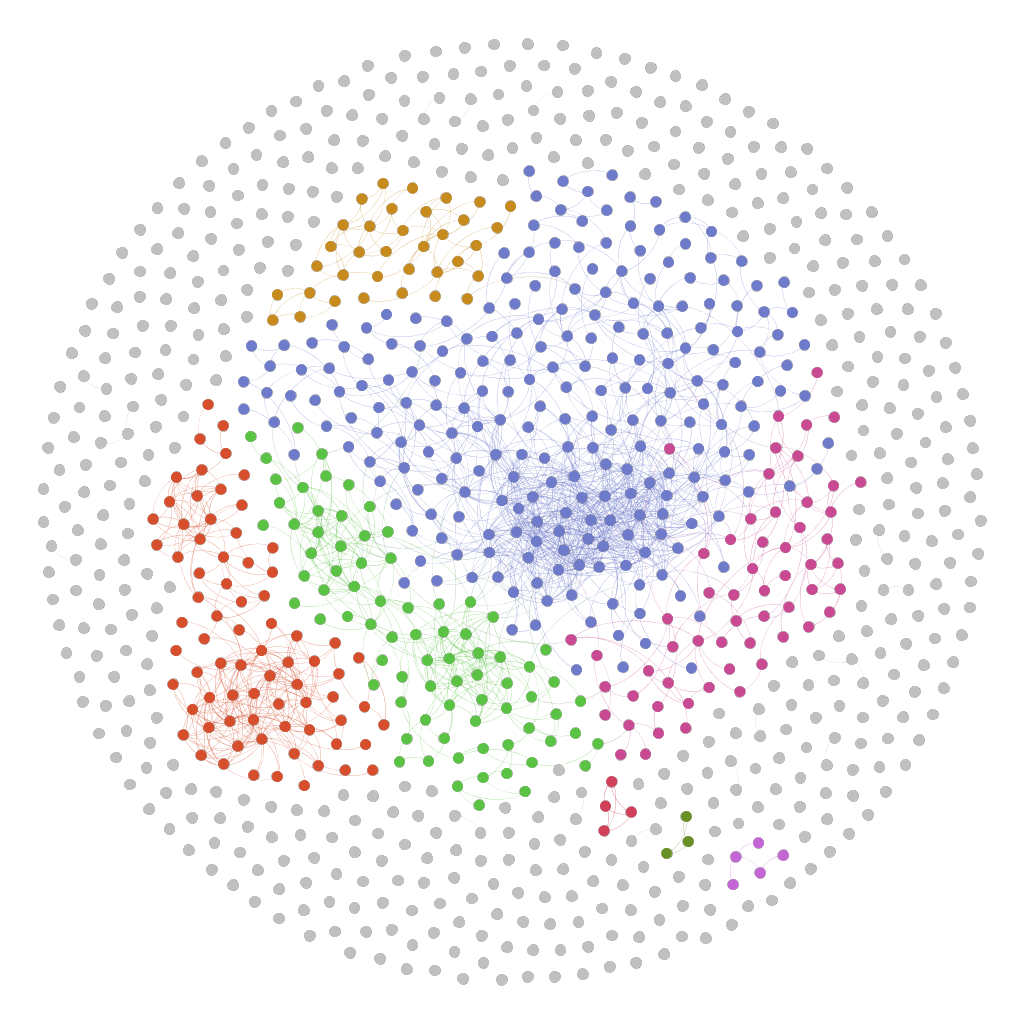}

\caption{Thresholding correlation networks. The researcher must solve a model selection problem when thresholding a correlation network, here exemplified by gene co-expression data from the plant {\it Arabidopsis thaliana}. As the threshold increases, more communities appear, potentially leading to overfitting (left). Including too many links (low threshold) can lead to underfitting (middle). Our module-based cross-validation method guides the researcher to the most parsimonious model (right).}
\label{fig:modelsel}
\end{figure}

Network communities -- groups of densely connected nodes -- are one of the most relevant network structures. They play a crucial role in the function of disparate complex systems, from metabolic networks \cite{guimera2005functional} to ecological communities \cite{calatayud2020positive}. When considering the community structure, selecting a threshold in a correlation network presents the researcher with a model selection problem with the usual pitfalls of overfitting and underfitting. When increasing the threshold of which links to include, the network becomes sparser and more communities appear (Fig.\ \ref{fig:modelsel}). If only strong links remain, the community structure can be rich and potentially informative about the modular structure of the underlying data. However, removing too many links leads to overfitting since spurious communities appear in the overly sparse network \cite{smiljanic1}. Including too many weak and noisy links, on the other hand, leads to underfitting, resulting in a dense network with few or no communities (see Fig.\ \ref{fig:modelsel}). The researcher must then find the best balance between underfitting and overfitting using standard procedures such as cross-validation. In the present work we propose a cross-validation method that focuses on communities, where the model complexity corresponds to the modular structure, in this way integrating thresholding and network community structure into a single step. 

We use the modular compression as quantified by the community-detection objective function known as the map equation \cite{RosvallPNAS2008} combined with cross-validation to find the threshold that gives the best balance between over- and underfitting network communities. We illustrate the proposed module-based cross-validation method using synthetic data and gene co-expression data of the plant {\it Arabidopsis thaliana}. We also use the code length savings measured by the map equation to assess the number of samples needed to discern structure in experimental data. We show that the code length savings plateau with a step-like relation to the number of samples. This behaviour is essential to the practitioner since the reward for sampling effort varies strongly with the number of samples. We also compare our approach to the widely used WGCNA method and see that this method underfits the model to the data, possibly failing to identify relevant network structure.


\section*{Results}
A correlation network is constructed from data having the same structure across scientific disciplines. Denoting the data $X\in \mathbb{R}^{p\times n}$, we have $n$ nodes or features that vary across $p$ experiments, samples or time instances. We construct a correlation network $\mathcal{G}(\hat{\Sigma})$ from the empirical correlation matrix $\hat{\Sigma}$ whose elements $\hat{\Sigma}_{i,j} = |\rho(X_i, X_j)|$ are given by the absolute value of the correlation $\rho$ between measurements at node $i$ and $j$. For a threshold $\tau$, the correlation matrix elements are given by
\begin{equation}
    \hat{\Sigma}_{i,j}^{\tau} = 
    \begin{cases}
        |\rho (X_i, X_j)|, & |\rho (X_i, X_j)| \geq \tau \\
        0, & |\rho (X_i, X_j)| < \tau.
    \end{cases}
\end{equation}


\subsection*{Module-based cross-validation}

To find the best balance between over- and underfitting in a correlation network, we propose a module-based cross-validation procedure. We split the data into a training and a test set $X^{train}\in \mathbb{R}^{p'\times n}$ and $X^{test}\in \mathbb{R}^{(p-p')\times n}$, and construct the corresponding networks $\mathcal{G}(\hat{\Sigma}^{\tau, train})$ and $\mathcal{G}(\hat{\Sigma}^{\tau, test})$ using a specific threshold $\tau$. To assess the support for a modular structure in the data, we use the map equation framework \cite{RosvallPNAS2008,Rosvall2}, which exploits the minimum description length principle: the modular structure with the shortest description best explains the data. The map equation measures the per-step average code length $L$ required to encode a random walk on a network with a given partition $M$.
The greedy optimisation algorithm Infomap \cite{Edler1} seeks the partition that minimises the code length and reveals the most modular structure of the network with respect to the random-walk process on the network. Denoting the code length for a given partition $M$ of the training network $\mathcal{G}(\hat{\Sigma}^{\tau, train})$ with threshold $\tau$ as $L^{\tau, train}(M)$, Infomap seeks to solve the optimisation problem
\begin{equation}
    M^{\tau, train} = \argmin_M L^{\tau, train}(M),
\end{equation}
where $M^{\tau, train}$ is the optimal partition of the training network with threshold $\tau$. This partition is a model of the modular structure in the training data, and the map equation framework can quantify how well this model fits the test data.

If the modular structure in the training network is present in the test network, $M^{\tau, train}$ will also compress the modular description of the test network, decreasing the code length when applied to the test network. We quantify this compression by evaluating the relative code length savings, which we denote $l$ and define as 
\begin{equation}
    l^\tau = \frac{L^{test}(1)-L^{test}(M^{\tau, train})}{L^{test}(1)} 
\end{equation}
for a particular value of the threshold $\tau$. $L^{test}(1)$ is the code length of the one-level uncompressed partition with all nodes in the same module. If $l^\tau>0$, the training modules are present also in the test network. The optimal threshold choice $\tau^*$ maximises the relative code length savings. Equivalently, the optimal threshold minimises the code length for the test network when using the optimal partition of the training network,
\begin{equation}
    \tau^* = \argmax_\tau l^\tau = \argmin_\tau L^{test}(M^{\tau, train}).
    \label{eq:opt_problem}
\end{equation}
The partition $M^{\tau^*, train}$ for the optimal $\tau^*$ best generalises to the independent test data set, balancing over- and underfitting. This module-based cross-validation allows us to select the optimal threshold and identify reliable modules. 

\subsection*{Synthetic data}
To illustrate the method using synthetic data we sample from a multivariate normal distribution with a planted partition. The covariance matrix elements are
\begin{equation}
    \Sigma_{i,j} = 
    \begin{cases}
        1, & i=j \\
        c, & M(n_i)=M(n_j)\\
        0, & M(n_i)\neq M(n_j),
    \end{cases}
\end{equation}
where we use $M(n_i)$ to denote the module of node $n_i$. The covariance matrix is block diagonal with the blocks corresponding to the planted partition with within-module covariance $c\in (0, 1]$, ensuring that the matrix is positive definite. The empirical correlation matrix $\hat{\Sigma}$ can then be calculated from samples $X \sim N(0, \Sigma)$. In the following we use Spearman's rank correlation, but other correlation measures can be used as well.

\begin{figure}[tb]
\includegraphics[scale=0.45]{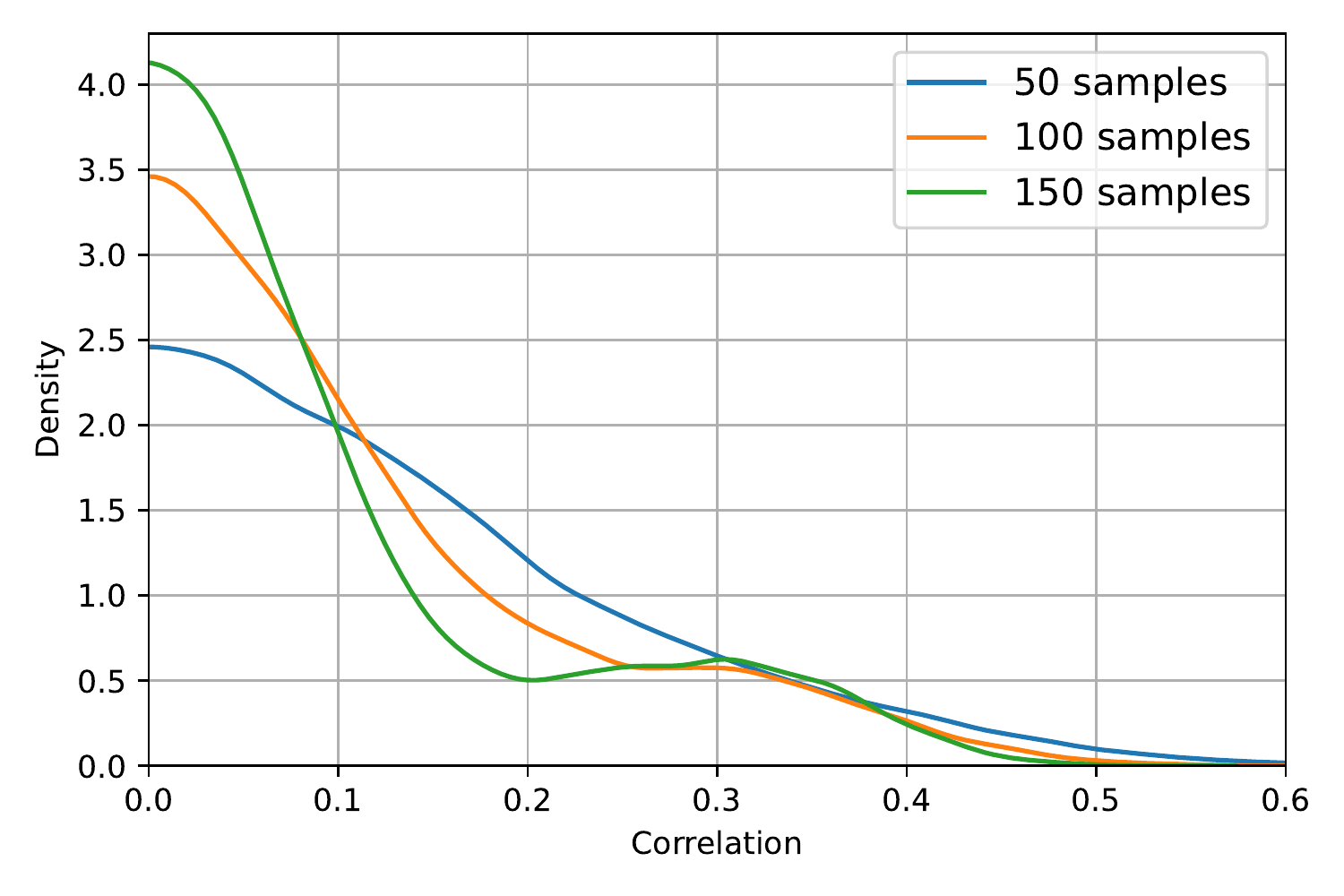}
\includegraphics[scale=0.45]{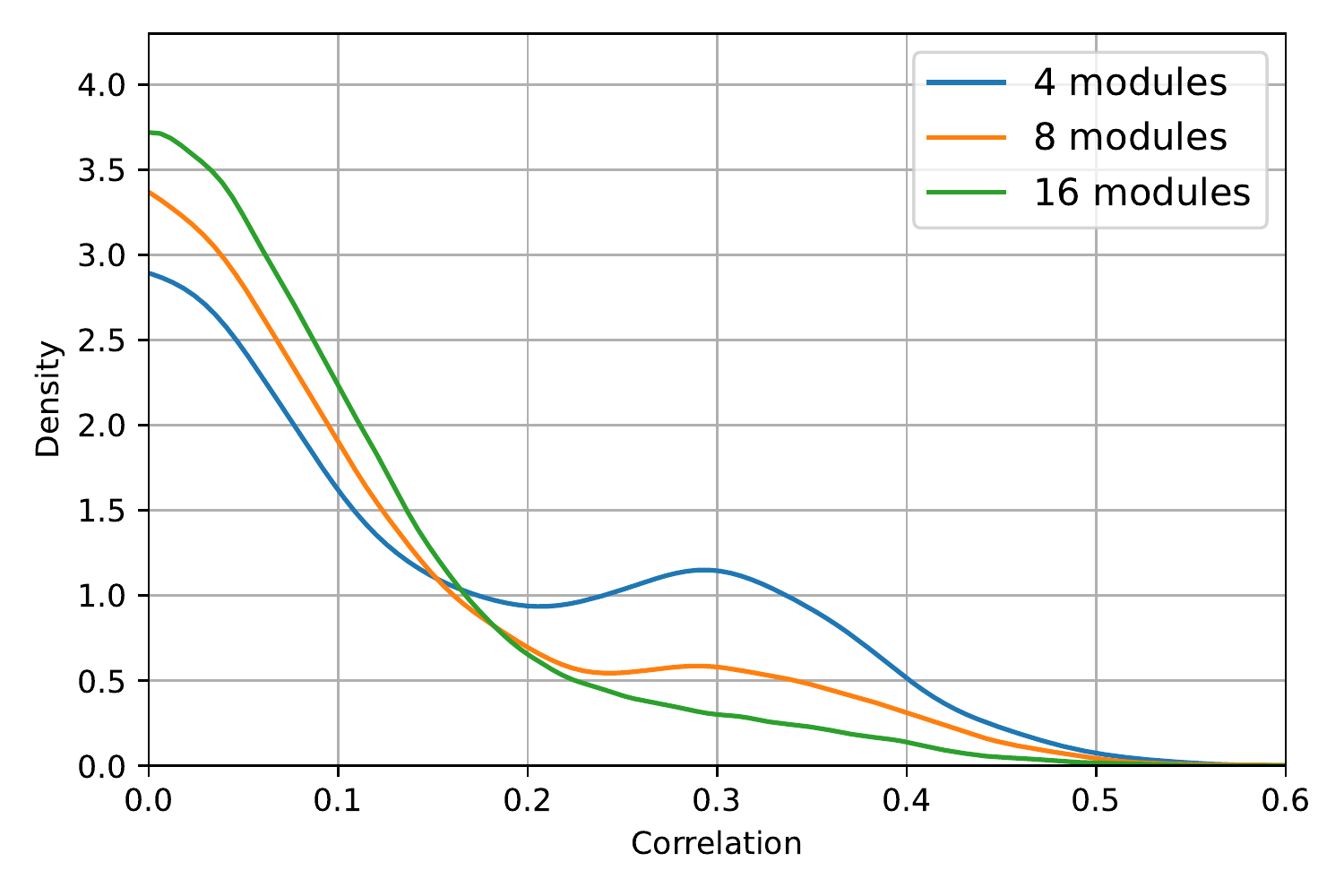}
\includegraphics[scale=0.45]{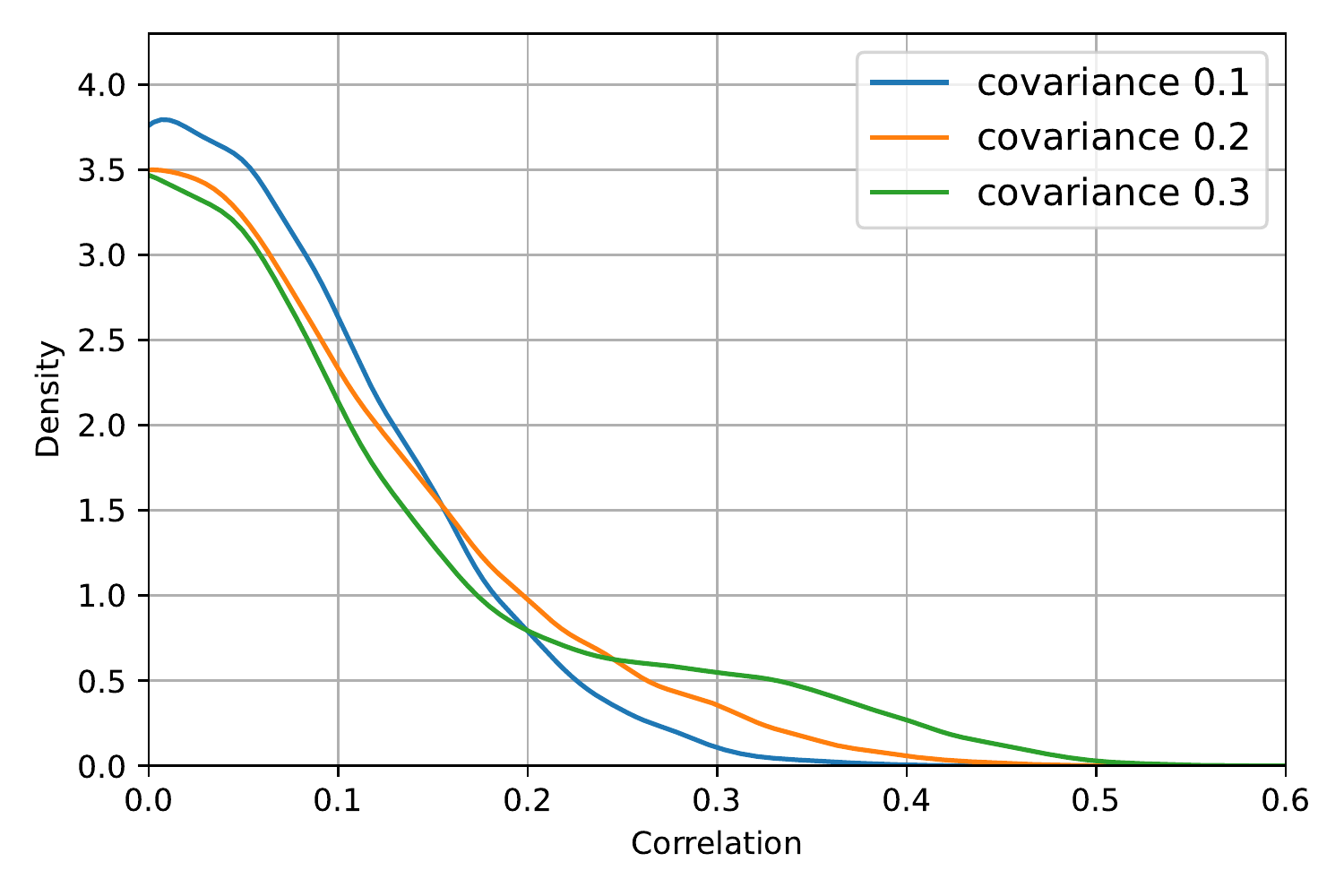}
\includegraphics[scale=0.45]{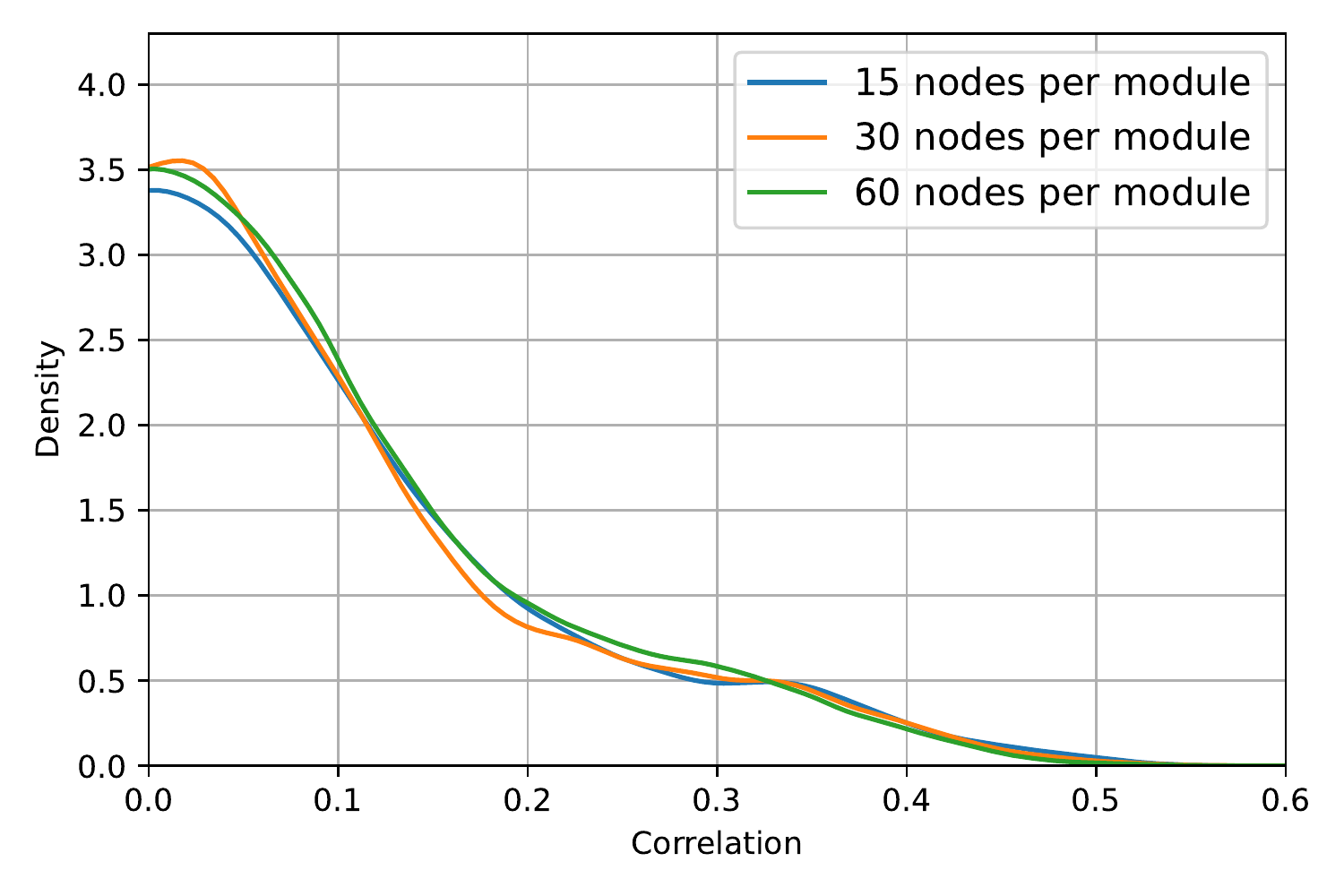}
\caption{Synthetic data. The distribution of empirical correlations calculated from the planted partitions varies with the number of samples (upper left), number of modules (upper right) and planted covariance (lower left), but seemingly not with the module size (lower right). More samples, fewer modules and large within-module covariance gives a stronger signal of modular structure in the data.}
\label{fig:synth_data_ex}
\end{figure}

The ability to separate signal from noise varies with the parameters in the sampled data. We therefore establish a baseline case with 8 modules with 30 nodes in each module, 100 samples and covariance $c=0.3$, and explore how the distribution of empirical correlations changes when we vary these parameters. With increasing number of samples the planted correlations are more easily distinguished from noise-level correlations (Fig.\ \ref{fig:synth_data_ex}). The number of modules has an opposite effect, with a clear peak of planted correlations with few modules. This can be explained considering that the fraction of within-module links and total links decreases as the inverse of the number of modules, meaning that the within-module links cover a larger part of the covariance matrix with fewer modules, thus increasing the signal-to-noise ratio. The number of nodes per module has little effect on the distribution of empirical correlations, because adding more nodes to a module increases the support for the corresponding correlation in the data. And if the support is already sufficient, adding more nodes will not change the distribution. Having only small modules can, however, change the distribution, and increases the sensitivity to stochastic variations. Hence, there is no simple relation between the number of samples and the number of nodes, but more nodes does not necessarily require more samples to recover a planted partition. Obviously, the planted within-module covariance $c$ affects the distribution, with a higher value leading to the planted correlations being farther away from the noise-level. This result shows the importance of including varying conditions in experiments.

\begin{figure}[tb]
\includegraphics[scale=0.5]{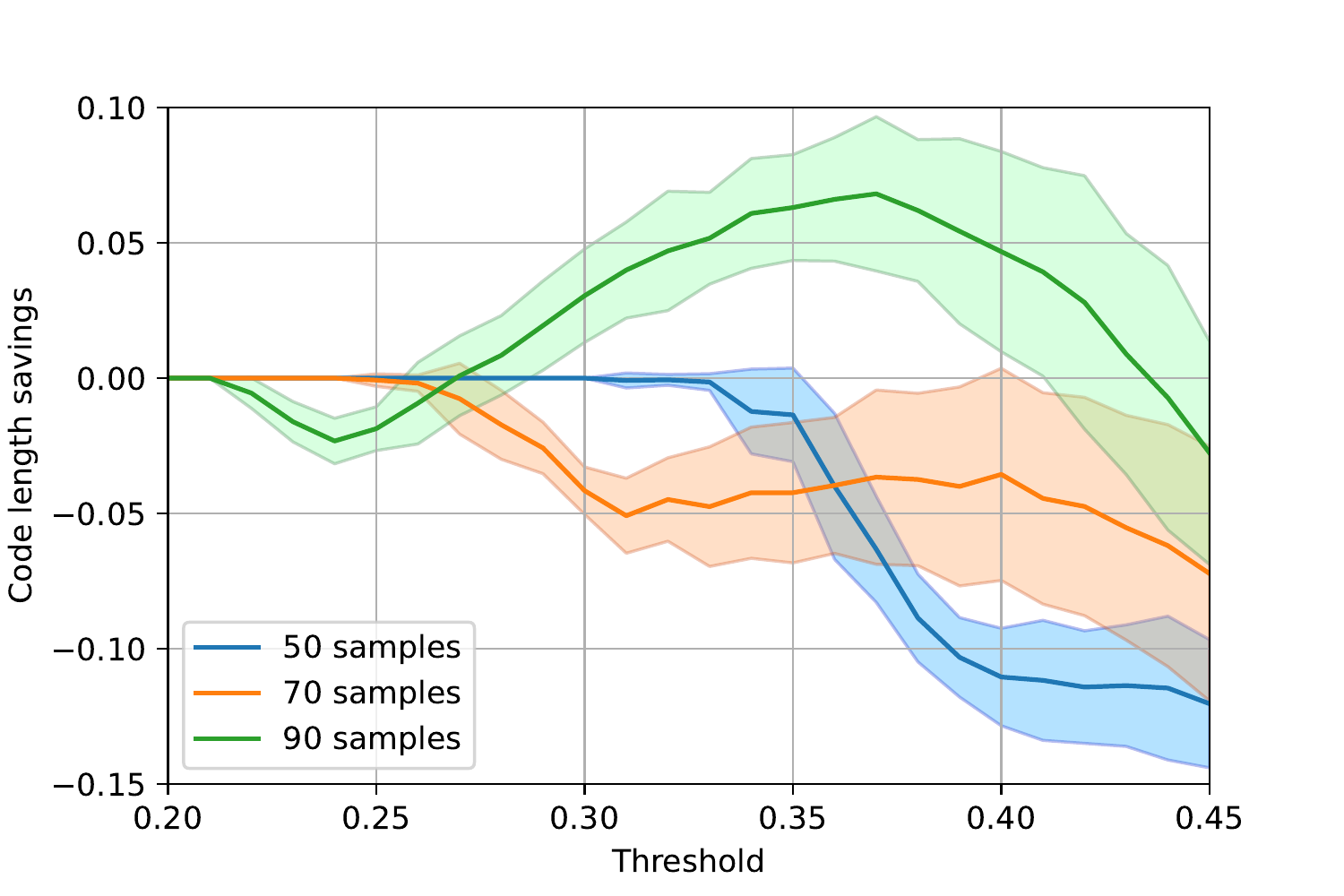}
\caption{Code length savings with synthetic data. The code length savings $l$ as a function of the threshold $\tau$ for different sample sizes and planted covariance $c=0.3$. With enough samples we get a positive code length saving in the test network, but with fewer samples there is no signal of the modules in the train network. The code length savings are negative when there are spurious modules, or zero when there is only one module. Calculations are averaged over ten runs and the shaded areas show the standard deviation.}
\label{fig:obj_func}
\end{figure}

Among the four variables expected to affect the signal-to-noise ratio, the practitioner can often control the number of samples, but there are often economic and other constraints that limit the availability of samples. Based on this, and given the analysis above showing the importance of the number of samples, we focus on how our proposed method performs when varying the number of samples, and how this affects the possibility to recover a planted partition. Figure \ref{fig:obj_func} shows the code length savings $l$ for our baseline case with varying number of samples. With 90 samples, $l$ peaks for a specific value of the threshold, corresponding to the solution $\tau^*$ of the optimisation problem in Eq.\ \ref{eq:opt_problem}. With fewer samples, however, the code length savings never exceed zero, irrespective of the threshold. The intervals with zero code length savings correspond to a dense network with no modular structure. The optimisation problem must therefore be stated as
\begin{equation}
    \tau^* = \argmax_\tau l^\tau, \quad \mathrm{s.t.} \quad l^\tau>0, \quad m^\tau>1,
\end{equation}
where $m^\tau$ denotes the number of modules. The added constraints only allow solutions with a modular structure present in both the training and test networks. Given the data with 50 and 70 samples in Fig.\ \ref{fig:obj_func}, we do not find a solution: a proper threshold without over- or underfitting is not feasible with this method. These results show the need for sufficient sampling to reliably detect structure in the underlying data. More importantly, the results also suggest that the proposed method can be used to estimate a sufficient number of samples. 

\begin{figure}[tb]
\includegraphics[scale=0.5]{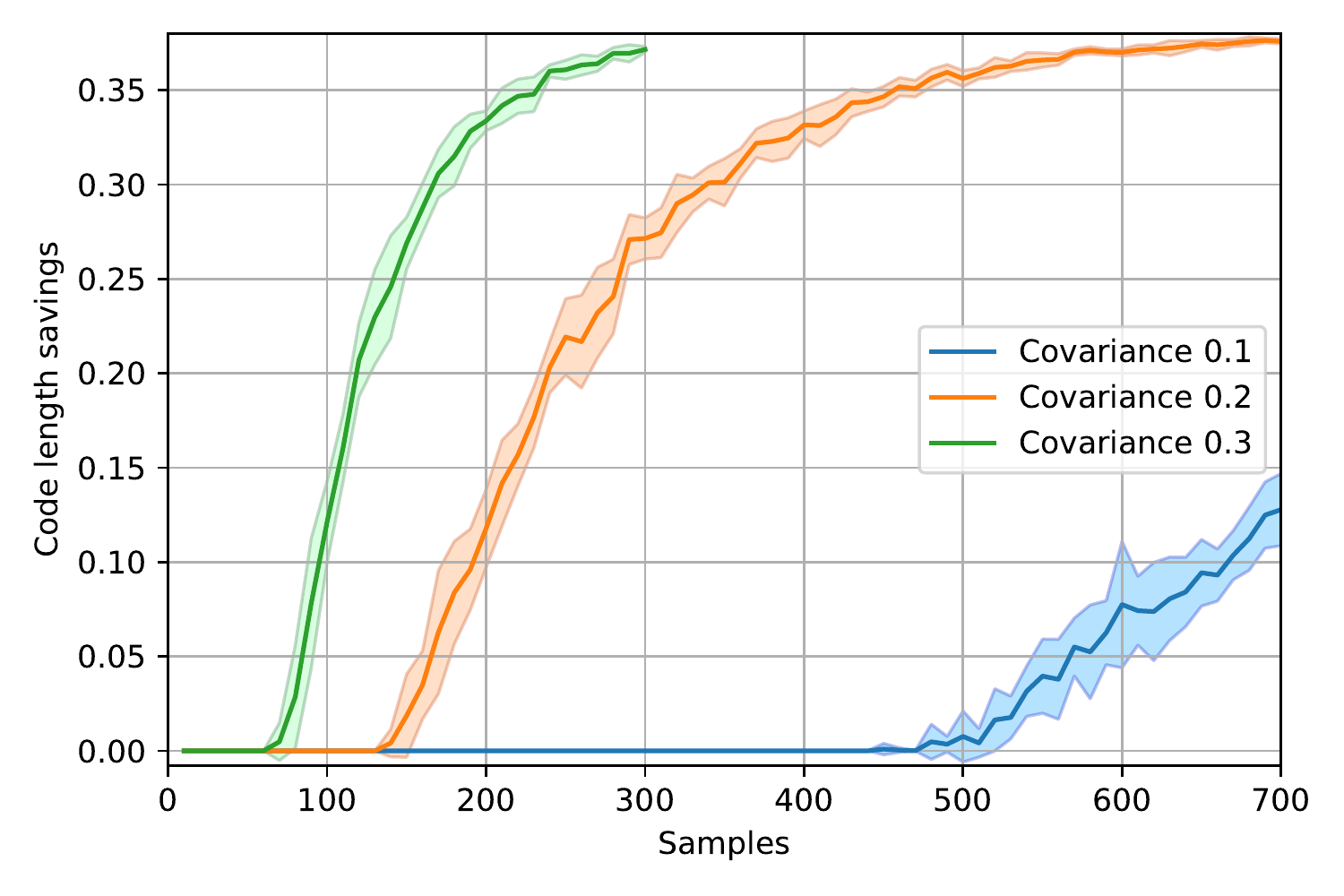}
\caption{Code length savings with varying number of samples. The code length savings for different values of the planted covariance increase as the number of samples increases. For the larger covariances here, the increase is almost step-like and the reward for increased sampling effort varies strongly.}
\label{fig:CLsav_synth}
\end{figure}

\begin{figure}[tb]
\includegraphics[scale=0.5]{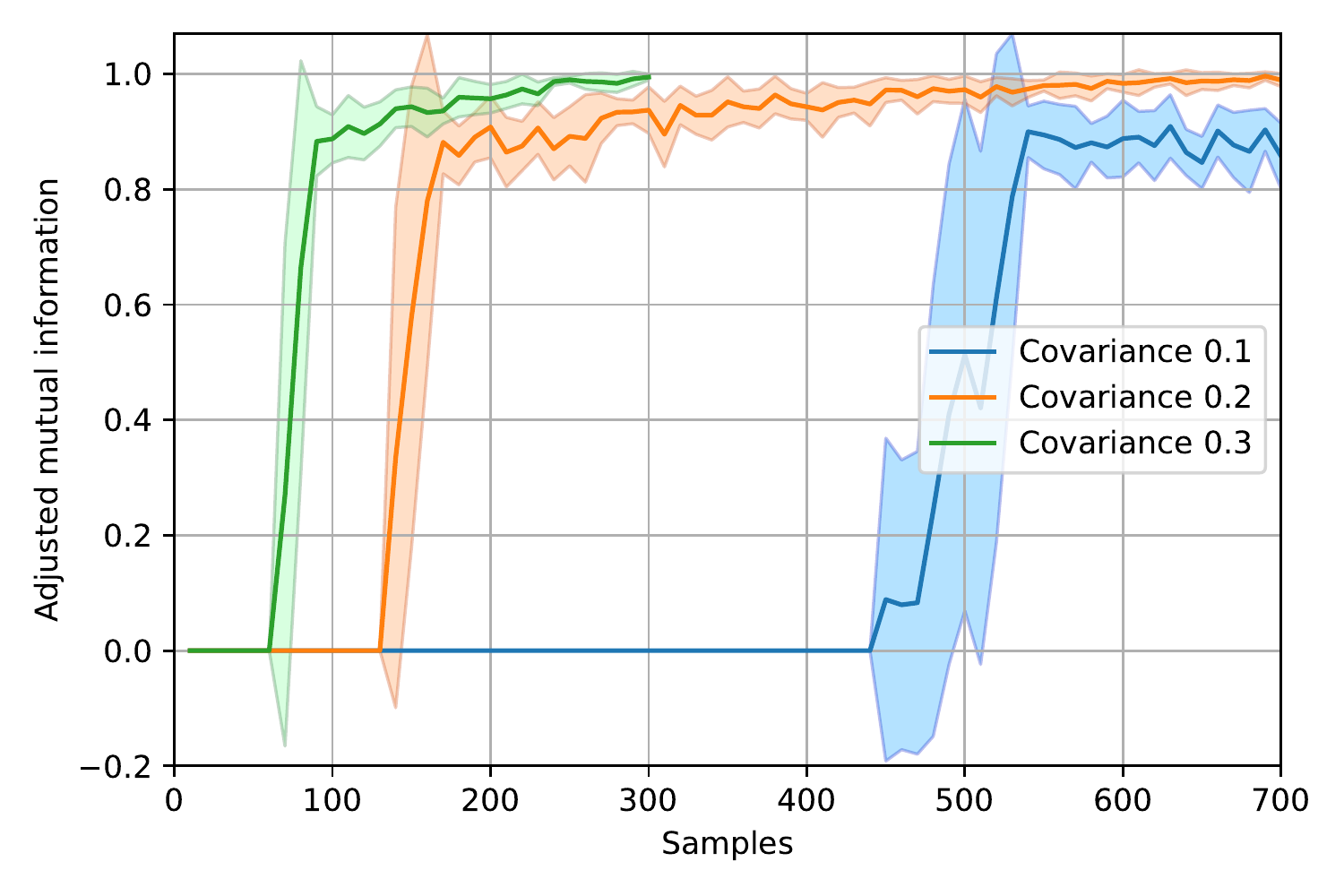}
\caption{AMI with varying number of samples. The adjusted mutual information (AMI) between planted and recovered partitions increases rapidly and is large ($>0.8$) for a wide range of samples. Further increasing the number of samples only gives small gains in AMI.}
\label{fig:AMI_synth}
\end{figure}

To further investigate the role of the number of samples, we calculate the code length savings for the optimal threshold $\tau^*$ when varying the number of samples for different values of the within-module covariance $c$. The code length savings increase rapidly when there are enough samples to obtain a signal of the modules in the train network also in the test network (Fig. \ref{fig:CLsav_synth}). When this increase happens and how sharp the increase is depends on the within-module covariance, with a covariance $c=0.1$ requiring many samples. The code length savings flatten out as the number of samples is increased further, indicating that there is less reward, in terms of finding modular structure, after a certain point. This means that the method we propose can be used to estimate an optimal sampling effort, defined as the number of samples where the code length savings start to plateau. Figure \ref{fig:AMI_synth} illustrates this effect further: For a large enough number of samples we recover the planted partition fairly well, with adjusted mutual information (AMI) larger than 0.8. Further increasing the number if samples only gives small gains in AMI. 
These results resemble those of Decelle et al.\ \cite{Decelle:PhysRevLett} that identify a phase transition in the detection of partitions planted using the stochastic block model when going from a random graph to separated modules, which is their way of increasing the signal-to-noise ratio. Our results are also based on modular structure and show that the detection of planted partitions undergo a process resembling a phase transition when increasing the number of samples. 

\subsection*{Gene co-expression data}

\begin{figure}[tb]
\includegraphics[scale=0.37]{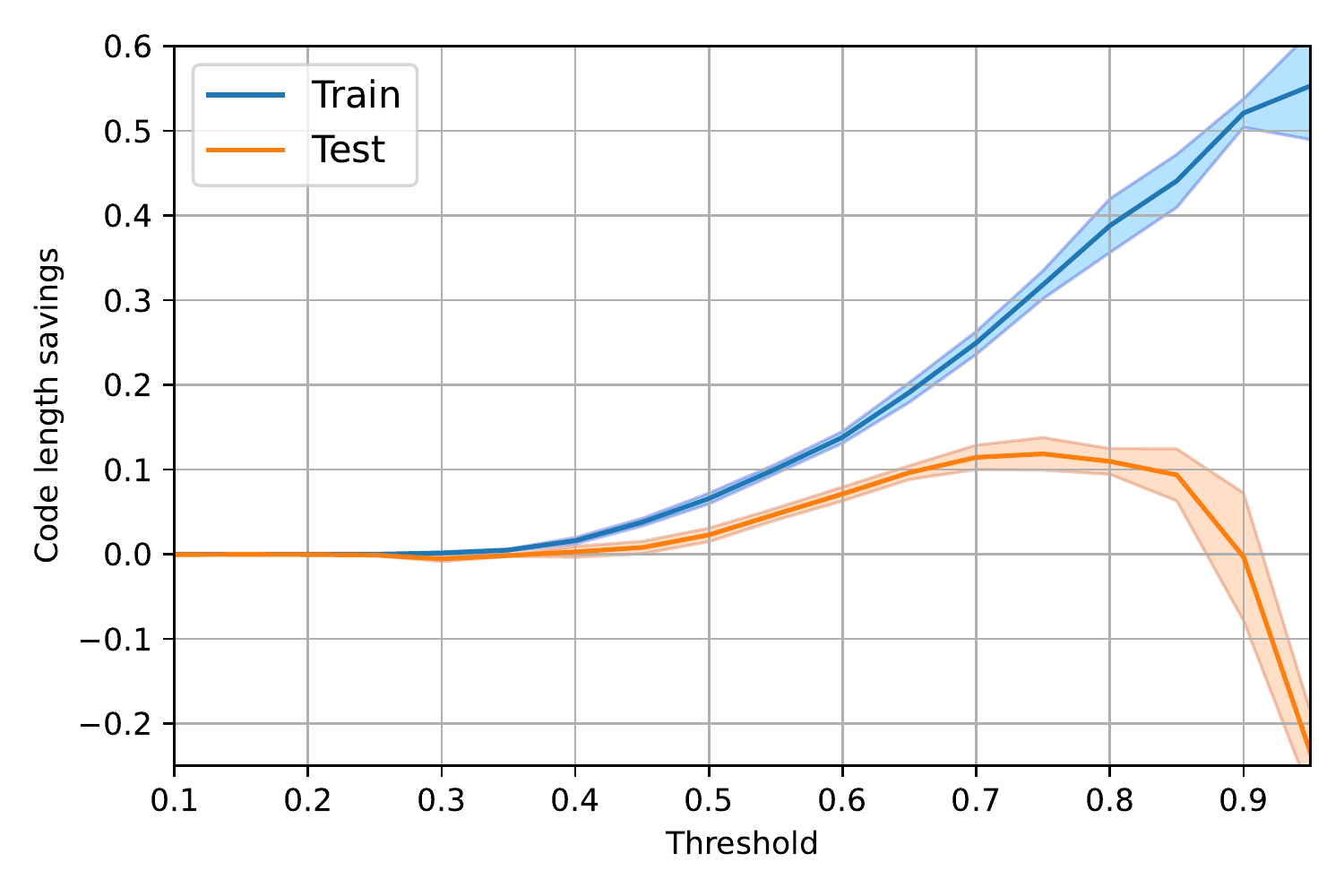}
\includegraphics[scale=0.37]{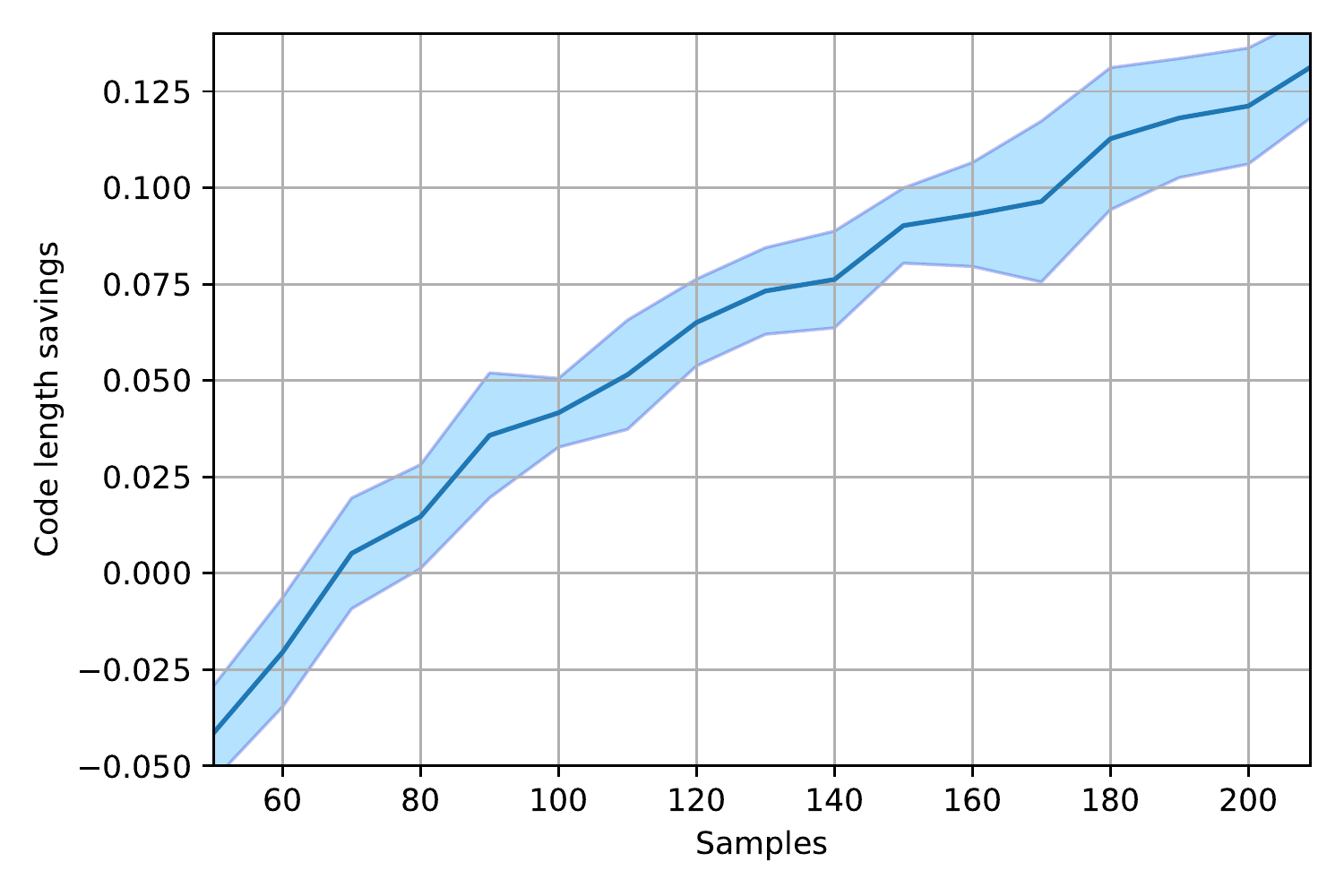}
\caption{Gene co-expression data. Applying the proposed method to gene co-expression data from the plant {\it Arabidopsis thaliana} shows that the code length savings in the test network has a maximum at threshold $\tau^* \approx 0.75$ (left). This is the threshold value giving the most parsimonious model of the data. The right figure shows that the code length savings at the optimal threshold do not increase linearly with the number of samples, indicating in accordance with the results based on synthetic data that the reward for sampling effort decreases as the number of samples increases.}
\label{fig:crossval_genes}
\end{figure}

As an example using real data, we choose gene co-expression data from the plant {\it Arabidopsis thaliana} (see Methods for details). The objective in this application is to see how genes are co-expressed when the plant is subject to stress such as heat or cold. When we apply our proposed method to these data we see that the code length savings in the training network for different thresholds $\tau$ increase as the threshold increases (Fig. \ref{fig:crossval_genes}). This increase is to be expected since the network becomes sparser as the threshold increases, resulting in a larger modular compression of the network. Contrary to this, the code length savings $l^\tau$ in the test network increase up to a certain point $\tau^* \approx 0.75$ and then start to decrease. The optimal partition of the training network maximises the modular compression of the test network at $\tau^*$. This threshold gives the most parsimonious model of the data when considering the modular structure. The networks in Fig. \ref{fig:modelsel} are based on these data, with the right-most one corresponding to $\tau^* = 0.75$.

Figure \ref{fig:crossval_genes} also shows the maximum code length savings as the number of samples increases, using these same data but randomly undersampling the number of samples. The dependence on samples is non-linear, resembling the behaviour for synthetic data.

\begin{figure}[tb]
\includegraphics[scale=0.5]{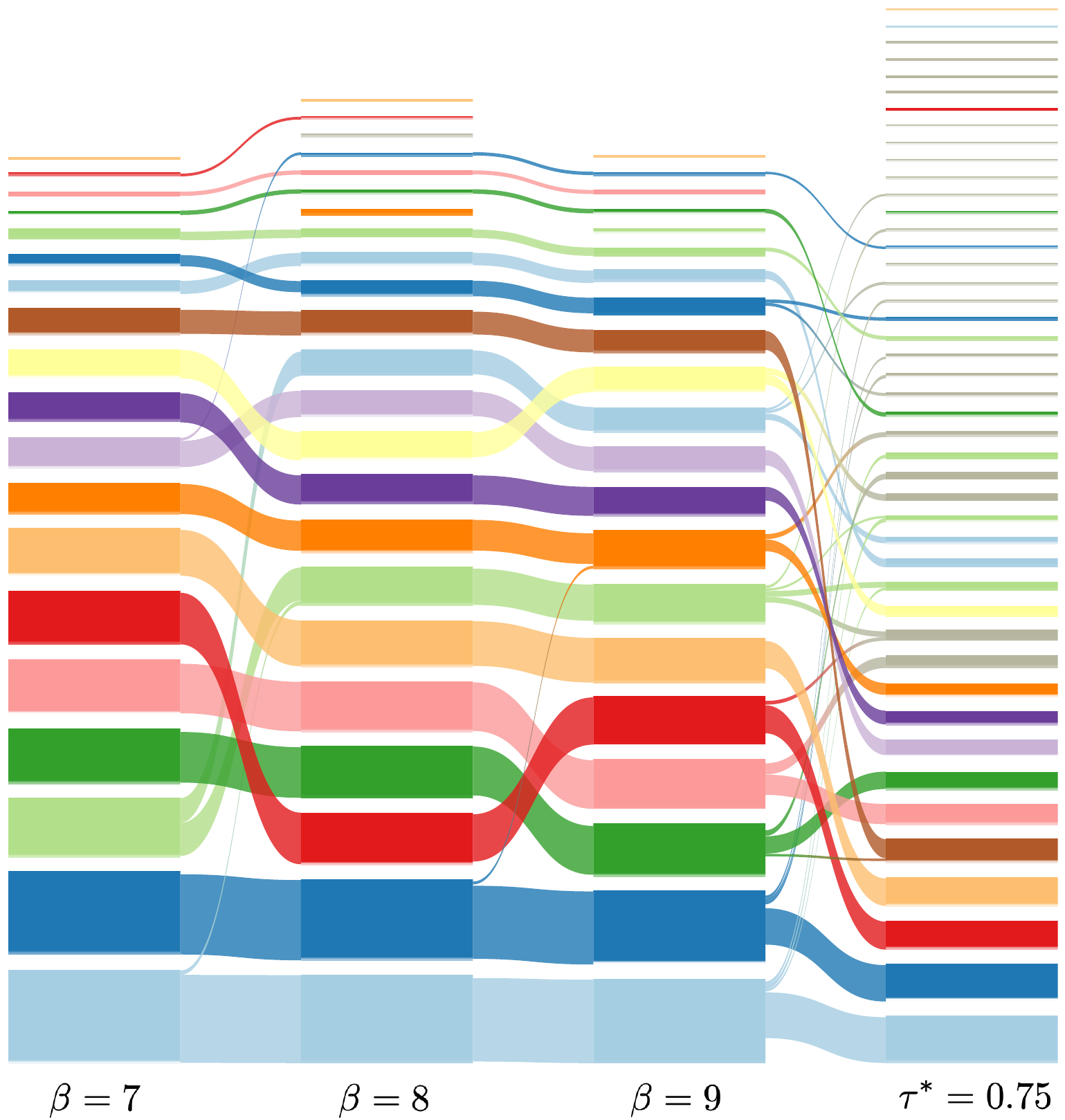}
\caption{WGCNA underfits modules. The allluvial diagram shows Infomap modules and reveals that the networks derived using WGCNA with different $\beta$ values give less modular structure than the network corresponding to threshold $\tau^*=0.75$. This shows that WGCNA underfits the modular structure in the data.}
\label{fig:wgcna}
\end{figure}

Gene co-expression data are often analysed using weighted gene co-expression network analysis (WGCNA) \cite{Horvath}. WGCNA employs soft thresholding by introducing a parameter $\beta$ such that the network link weights are $|\rho|^\beta$ for a correlation measure $\rho$. We use WGCNA to assess whether this method over- or underfits the modular structure studied in this work, and find that WGCNA underfits the model to the data. For computational reasons, we select the 4000 genes with the largest variance and find that $\beta = 8$ is a good choice (see Methods for details). Figure \ref{fig:wgcna} shows an alluvial diagram with network modules obtained using Infomap for networks corresponding to different $\beta$ values and for the network corresponding to the threshold $\tau^*$. The threshold $\tau^*$ that best balances between over- and underfitting gives a more modular network than WGCNA, meaning that WGCNA underfits in this sense.

\section*{Conclusions}


We have suggested a way of solving the common and general problem of thresholding correlation networks by integrating modular structure in the model selection process. This approach avoids ad-hoc methods to sparsify networks that risk over- or underfitting modular structure. Our method quantifies the partition quality by means of the code length in the map equation, which we use to measure how a learned community structure fits unseen data. Experiments on synthetic and real data show promising results. The applicability of the method depends on the signal-to-noise ratio, which we showed depends on a number of factors, including within-module covariance, module size and the number of samples. There is no obvious relation between the number of nodes and the number of samples, but we showed that more nodes does not necessarily require more samples.

The reward for sampling effort varies strongly as the number of samples increases, and the ability to recover a partition undergoes what resembles a phase transition. Given the ubiquity of the problem and the substantial economic cost of sampling, our results indicating that increased sampling effort can have little reward could potentially lead to reduced spending and better allocation of resources across scientific fields working with correlations in multivariate data.

The comparison with WGCNA showed that there is a risk of underfitting when using WGCNA, which can lead to meaningful network structure being overlooked. 


\section*{Methods}
\subsection*{Optimization method}
The objective function in the optimization problem (\ref{eq:opt_problem}) that we solve is the relative code length savings in the test network, $L^{test}(M^{\tau,train})$, that depend on the threshold $\tau$. This objective function is stochastic for two reasons. The first and main reason is the two-fold splitting that we use here in the cross-validation. The second reason is the inherent randomness in Infomap and the non-convex properties of the map equation \cite{Calatayud}. To overcome this we do a sample average approximation and average over 10 runs, such that $L^{test}(M^{\tau,train}) = \frac{1}{10}\sum_{i=1}^{10} L^{test_i}(M^{\tau,train_i})$. In this work we use a simple approach to find the optimal $\tau$ by testing a set of thresholds, such as $\tau \in [0.1, 0.2, \ldots, 0.9]$, and choosing the one that maximizes the code length savings and thus minimizes $L^{test}(M^{\tau,train})$. Regarding the two-fold splitting, we believe that this is the current best option to build both a training and a test network from a data set. There are however recent developments in regularization methods for link prediction \cite{smiljanic1} that can potentially provide better options.

\subsection*{Gene expression data and analysis}

Gene expression data for the co-expression network were retrieved from the Sequence Read Archive (SRA). We searched SRA to identify all available RNA-Seq samples relating to cold stress in {\it Arabidopsis thaliana} ecotype Columbia-0, but limited ourselves to leaf tissue and excluded any genetic variants. The selected data sets include both control and treated samples and were retrieved in April 2021. A full list of included samples can be found in the supplementary material. The data were quantified using salmon version 1.2.1 \cite{patro2017salmon} against the Araport 11 release of the {\it Arabidopsis thaliana} genome. Pre-processing and normalization were done in R using the variance stabilizing transform available in DESeq2 \cite{love2014moderated}. If the expression value is constant across samples for a gene, due to the sub sampling of experiments, then the correlation coefficient for this gene and the corresponding network links are undefined, and the gene is disconnected from the co-expression network.

\begin{figure}[tb]
\includegraphics[scale=0.5]{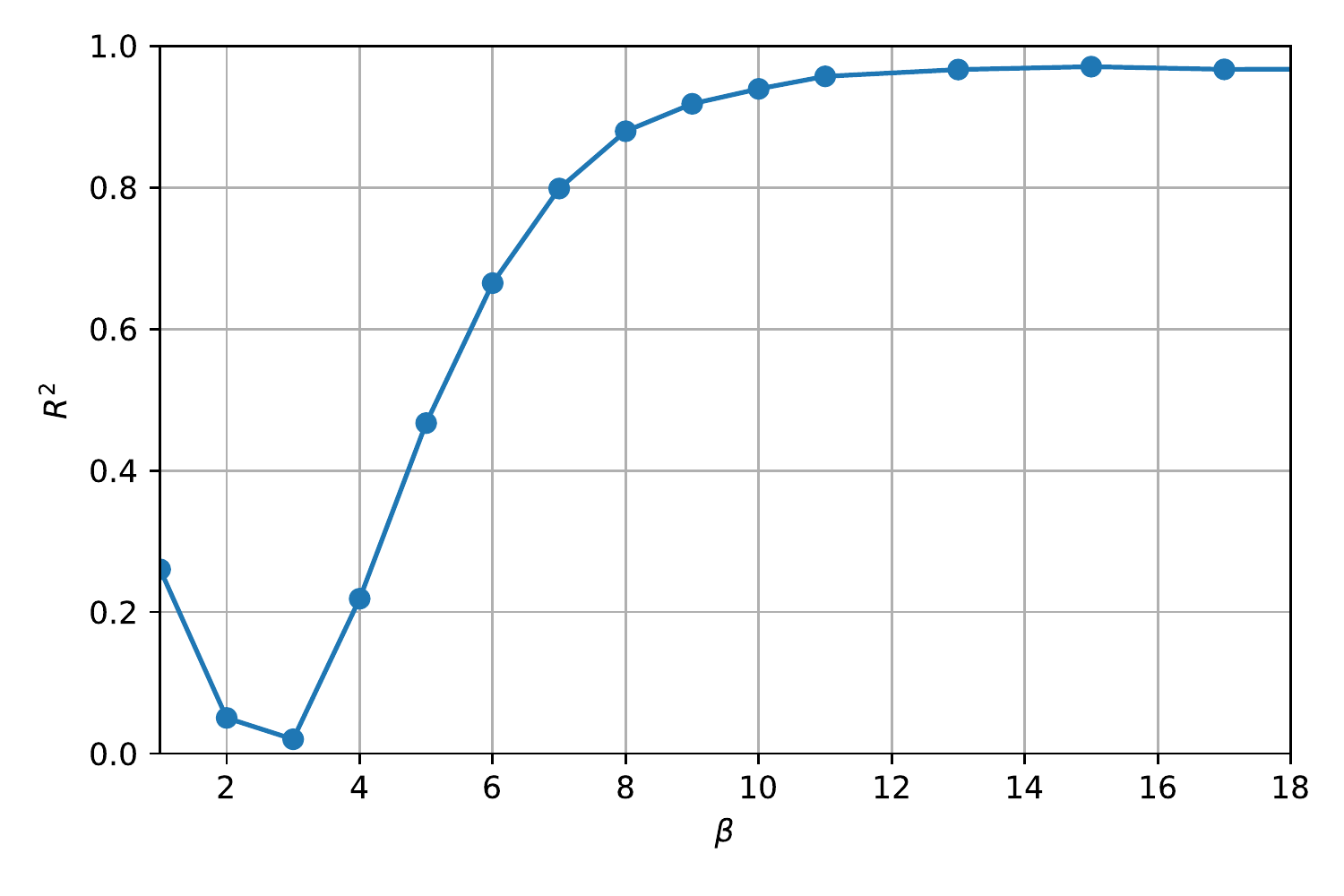}
\caption{WGCNA parameter. In WGCNA, a coefficient of determination $R^2$ larger than 0.8 indicates that the network is approximately scale-free, making $\beta = 8$ a good choice for our data.}
\label{fig:wgcna_beta}
\end{figure}

In WGCNA the parameter $\beta$ is chosen such that the co-expression network is approximately scale-free. This is done through linear regression of the distribution of weighted node degree on a log-log scale, where the smallest $\beta$ such that the coefficient of determination $R^2 > 0.8$ is chosen. Figure \ref{fig:wgcna_beta} shows the dependence of $R^2$ on $\beta$, indicating that $\beta = 8$ is a good choice for our data.

\section*{Declarations}
\subsection*{Availability of data and materials}
Gene expression data were retrieved from the Sequence Read Archive (SRA), https://www.ncbi.nlm.nih.gov/sra. A list of samples extracted from SRA can be found in the supplementary material.
\subsection*{Competing interests}
The authors declare that they have no competing interests.
\subsection*{Funding}
The computations were enabled by resources provided by the Swedish National Infrastructure for Computing (SNIC) at HPC2N partially funded by the Swedish Research Council through grant agreement no. 2018-05973.
\subsection*{Authors' contributions}
MN, VJ, and MR conceived and designed the study. MN performed the experiments. MN, VJ, JC and MR interpreted the results and wrote the paper. All authors read and approved the final manuscript.
\subsection*{Acknowledgements}
We thank A. Rojas, J. Smiljani\'c and A. Vergara for providing comments that improved the manuscript.


\begin{thebibliography}{2}
\bibitem{Wang} Wang, Y.X.R., Huang, H.: Review on statistical methods for gene net- work reconstruction using expression data. J Theor Biol 362, 53–61 (2014). doi:10.1016/j.jtbi.2014.03.040
\bibitem{Marbach} Marbach, D., Costello, J.C., Küffner, R., Vega, N.M., Prill, R.J., Camacho, D.M., Al- lison, K.R., Aderhold, A., Bonneau, R., Chen, Y., Collins, J.J., Cordero, F., Crane, M., Dondelinger, F., Drton, M., Esposito, R., Foygel, R., de la Fuente, A., Gertheiss, J., Geurts, P., Greenfield, A., Grzegorczyk, M., Haury, A.-C., Holmes, B., Hothorn, T., Husmeier, D., Huynh-Thu, V.A., Irrthum, A., Kellis, M., Karlebach, G., Lèbre, S., De Leo, V., Madar, A., Mani, S., Mordelet, F., Ostrer, H., Ouyang, Z., Pandya, R., Petri, T., Pinna, A., Poultney, C.S., Rezny, S., Ruskin, H.J., Saeys, Y., Shamir, R., Sîrbu, A., Song, M., Soranzo, N., Statnikov, A., Stolovitzky, G., Vega, N., Vera-Licona, P., Vert, J.-P., Visconti, A., Wang, H., Wehenkel, L., Windhager, L., Zhang, Y., Zim- mer, R., Consortium, T.D.: Wisdom of crowds for robust gene network inference. Nature Methods 9(8), 796–804 (2012). doi:10.1038/nmeth.2016
\bibitem{Bullmore} Bullmore, E., Sporns, O.: Complex brain networks: graph theoretical analysis of struc- tural and functional systems. Nature Reviews Neuroscience 10(3), 186–198 (2009). doi:10.1038/nrn2575
\bibitem{Barberan} Barberán, A., Bates, S.T., Casamayor, E.O., Fierer, N.: Using network analysis to explore co-occurrence patterns in soil microbial communities. The ISME Journal 6(2), 343–351 (2012). doi:10.1038/ismej.2011.119
\bibitem{Horvath} Zhang, B., Horvath, S.: A general framework for weighted gene co-expression network analysis. Stat Appl Genet Mol Biol 4, 17 (2005). doi:10.2202/1544-6115.1128
\bibitem{deVries} de Vries, F.T., Griffiths, R.I., Bailey, M., Craig, H., Girlanda, M., Gweon, H.S., Hallin, S., Kaisermann, A., Keith, A.M., Kretzschmar, M., Lemanceau, P., Lumini, E., Mason, K.E., Oliver, A., Ostle, N., Prosser, J.I., Thion, C., Thomson, B., Bardgett, R.D.: Soil bacterial networks are less stable under drought than fungal networks. Nature Commu- nications 9(1), 3033 (2018). doi:10.1038/s41467-018-05516-7
\bibitem{Civier} Civier, O., Smith, R.E., Yeh, C.-H., Connelly, A., Calamante, F.: Is removal of weak connections necessary for graph-theoretical analysis of dense weighted structural connectomes from diffusion mri? NeuroImage 194, 68–81 (2019). doi:10.1016/j.neuroimage.2019.02.039
\bibitem{Serrano} Serrano, M.Á., Boguñá, M., Vespignani, A.: Extracting the multiscale backbone of complex weighted networks. Proceedings of the National Academy of Sciences 106(16), 6483–6488 (2009). doi:10.1073/pnas.0808904106. https://www.pnas.org/doi/pdf/10.1073/pnas.0808904106
\bibitem{Tumminello} Tumminello, M., Aste, T., Matteo, T.D., Mantegna, R.N.: A tool for filter- ing information in complex systems. Proceedings of the National Academy
of Sciences 102(30), 10421–10426 (2005). doi:10.1073/pnas.0500298102. https://www.pnas.org/doi/pdf/10.1073/pnas.0500298102
\bibitem{Dianati} Dianati, N.: Unwinding the hairball graph: Pruning algorithms for weighted complex networks. Phys. Rev. E 93, 012304 (2016). doi:10.1103/PhysRevE.93.012304
\bibitem{Friedman} Friedman, J., Hastie, T., Tibshirani, R.: Sparse inverse covariance estimation with the graphical lasso. Biostatistics 9(3), 432–441 (2008). doi:10.1093/biostatistics/kxm045
\bibitem{Meinshausen} Meinshausen, N., Bühlmann, P.: High-dimensional graphs and variable se- lection with the Lasso. The Annals of Statistics 34(3), 1436–1462 (2006). doi:10.1214/009053606000000281
\bibitem{guimera2005functional} Guimera, R., Nunes Amaral, L.A.: Functional cartography of complex metabolic net- works. nature 433(7028), 895–900 (2005)
\bibitem{calatayud2020positive} Calatayud, J., Andivia, E., Escudero, A., Melián, C.J., Bernardo-Madrid, R., Stoffel, M., Aponte, C., Medina, N.G., Molina-Venegas, R., Arnan, X., et al.: Positive associations among rare species and their persistence in ecological assemblages. Nature ecology \& evolution 4(1), 40–45 (2020)
\bibitem{smiljanic1} Smiljanić, J., Edler, D., Rosvall, M.: Mapping flows on sparse networks with missing links. Phys. Rev. E 102, 012302 (2020). doi:10.1103/PhysRevE.102.012302
\bibitem{RosvallPNAS2008} Rosvall, M., Bergstrom, C.T.: Maps of random walks on complex net- works reveal community structure. Proceedings of the National Academy
of Sciences 105(4), 1118–1123 (2008). doi:10.1073/pnas.0706851105. https://www.pnas.org/content/105/4/1118.full.pdf
\bibitem{Rosvall2} Rosvall, M., Axelsson, D., Bergstrom, C.T.: The map equation. The European Physical Journal Special Topics 178(1), 13–23 (2009). doi:10.1140/epjst/e2010-01179-1
\bibitem{Edler1} Edler, D., Bohlin, L., Rosvall, M.: Mapping higher-order network flows in memory and multilayer networks with infomap. Algorithms 10(112) (2017). doi:10.3390/a10040112
\bibitem{Decelle:PhysRevLett} Decelle, A., Krzakala, F., Moore, C., Zdeborová, L.: Inference and phase transitions in the detection of modules in sparse networks. Phys. Rev. Lett. 107, 065701 (2011). doi:10.1103/PhysRevLett.107.065701
\bibitem{Calatayud} Calatayud, J., Bernardo-Madrid, R., Neuman, M., Rojas, A., Rosvall, M.: Exploring the solution landscape enables more reliable network community detection. Phys. Rev. E 100, 052308 (2019). doi:10.1103/PhysRevE.100.052308
\bibitem{patro2017salmon} Patro, R., Duggal, G., Love, M.I., Irizarry, R.A., Kingsford, C.: Salmon provides fast and bias-aware quantification of transcript expression. Nature methods 14(4), 417–419 (2017)
\bibitem{love2014moderated} Love, M.I., Huber, W., Anders, S.: Moderated estimation of fold change and dispersion for rna-seq data with deseq2. Genome biology 15(12), 1–21 (2014)
\end{thebibliography}

\end{document}